# Mobility in Graphene Double Gate Field Effect Transistors

M.C. Lemme[*], T.J. Echtermeyer, M. Baus, B.N. Szafranek, J. Bolten, M. Schmidt, T. Wahlbrink and H. Kurz

*Advanced Microelectronic Center Aachen (AMICA), Otto-Blumenthal-Str. 25, 52074 Aachen, Germany*

**Abstract**

In this work, double-gated field effect transistors manufactured from monolayer graphene are investigated. Conventional top-down CMOS-compatible processes are applied except for graphene deposition by manual exfoliation. Carrier mobilities in single- and double-gated graphene field effect transistors are compared. Even in double-gated graphene FETs, the carrier mobility exceeds the universal mobility of silicon over nearly the entire measured range. At comparable dimensions, reported mobilities for ultra thin body silicon-on-insulator MOSFETs can not compete with graphene FET values.

## 1. Introduction

Scaling of transistor dimensions as described and dictated by Moore's law has generated astonishing innovation cycles in silicon technology over the last four decades [1][2]. As a result, CMOS technology today stands out as a fundamental technology that enables the global information society. Looking into the future, silicon technology is expected to remain the work horse for electronic applications for at least fifteen more years, but innovations no longer stem from pure geometrical scaling according to Moore's law. Instead, an era of material-based scaling has emerged, where novel materials must be introduced into the standard CMOS process to further reduce manufacturing cost, improve performance and/or reduce leakage power.

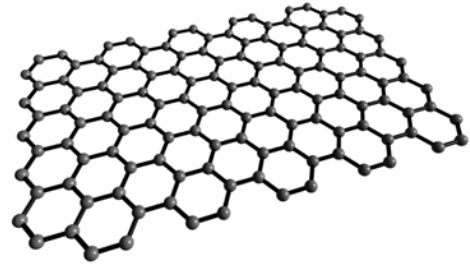

Fig. 1. Schematic of a graphene monolayer.

Carbon-based electronics are considered one of the most promising options to enhance silicon in the future [3]. A prominent example are carbon nanotubes (CNTs), which have received ample attention due to their intriguing electrical properties [4]. Nonetheless, there are two major drawbacks associated with CNTs. A lack of chirality control during production leads to a mixture of metallic and semiconducting nanotubes. In addition, there is no method currently available to accurately place hundreds of millions of nanotubes where they would be needed in order to form integrated circuits. Unless manufacturing processes are mastered and self-organization methods are employed, these could eventually prohibit their utilization as an addition to or replacement of silicon as a base material.

---
[*] corresponding author. Tel.: +49 241 8867 207; fax: +49 241 8867 571
E-mail address: lemme@amo.de (M.C. Lemme)

Another prominent example are carbon "substrates", thin carbon layers with similar properties to CNTs. Such two-dimensional single and few layer carbon sheets, referred to as graphene sheets, have only very recently been demonstrated to be thermodynamically stable [5]. A schematic of a graphene monolayer sheet is shown in Fig. 1: hexagonal rings formed by sp²-bonded carbon atoms are arranged in a dense honeycomb structure. In this two-dimensional form, graphene is a semi-metal with an extremely small overlap between the valence and the conduction band (zero-gap material). In its three-dimensional graphite form, graphene sheets are weakly coupled between the layers with van der Waals forces.

Carrier mobility values between 3000 and 27000 cm²/Vs have been reported for graphene and make it an extremely promising material for future nanoelectronic applications [5][6]. It is further known that carrier transport in graphene takes place in the π-orbitals perpendicular to the surface [7]. Intrinsically this translates into charge carrier transport with a mean free path for carriers of $L = 400$ nm at room temperature [5]. This would make ballistic devices feasible even at relaxed feature sizes compared to State-of-the-Art CMOS technology.

Published experimental data has been so far mainly obtained from mono- or few layer graphene on oxidized silicon wafers or decomposed intrinsic silicon carbide [5][6]. In first experiments, the surface of the graphene has been left uncovered and pseudo-MOS characteristics have been obtained. This is in contrast to any device integration, where a gate insulator, an electrode and low-k dielectrics need to be deposited on top of the graphene. Only very recently, first field effect devices with top-gate structures have been reported [8][9][10].

In this work, the electron and hole mobilities of graphene are extracted from field effect transistors (graphene FETs) with a double gate structure, where the pseudo-MOS back-gate is complemented by a second, lithographically defined sub-µm top-gate. A schematic of such a device is shown in Fig. 2. Carrier transport in these double-gated transistors is compared to "open" graphene FETs without top-gate dielectric. The extracted values are compared to literature data of silicon and ultra-thin body silicon on insulator (SOI) devices.

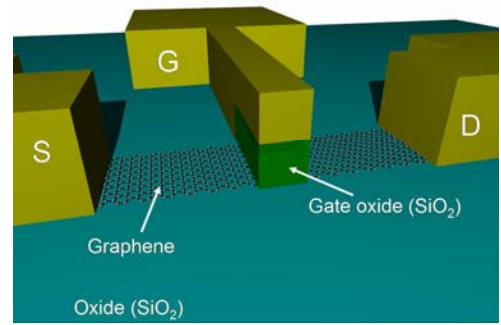

Fig. 2. Schematic of a graphene FET.

## 2. Experimental

High-quality silicon dioxide (SiO$_2$) films of $t_{ox}$ = 300 nm have been thermally grown on boron doped silicon wafers (100) with a base doping concentration of $N_A = 10^{15}$ cm$^{-3}$. A manual exfoliation process similar to that described in [5] has been used to deposit graphene on the oxidized wafers. Next, mono- and few layer graphene flakes have been visually identified with an optical microscope. Lift-off processes have been used to structure evaporated titanium (Ti) / gold (Au) source (S) and drain (D) contacts. At this point, IV-measurements of the pseudo-MOS structures have been obtained.

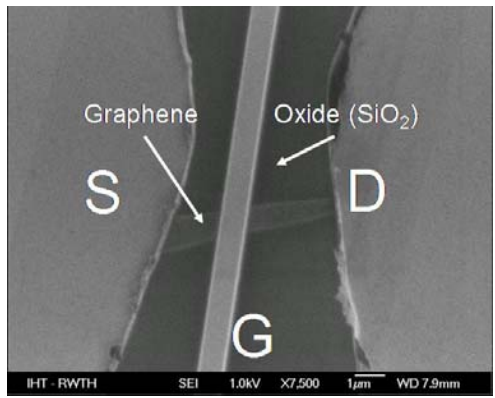

Fig. 3. Scanning electron microscope image of a graphene FET.

Next, electron beam lithography has been used to define top-gate electrodes. A gate stack of silicon dioxide gate dielectric (20 nm) and Ti (10 nm) / Au (100 nm) electrode has then been deposited by electron beam evaporation. Finally, the gate stack has been structured by a lift-off process. A scanning electron microscope image in Fig. 3 shows an example of a graphene field effect transistor. This particular graphene flake has a total length from source to drain of L = 5 µm and a width of W ≈ 650 nm under the top-gate. The e-beam defined top-gate transistor has a gate length of L = 650 nm.

Graphene thickness has been measured by atomic force microscopy, shown in Fig. 4. Here, the thickness of the graphene flake ranges from $t_g$ = 0.8 nm to $t_g$ = 2 nm. This indicates that only a few layers of graphene are present.

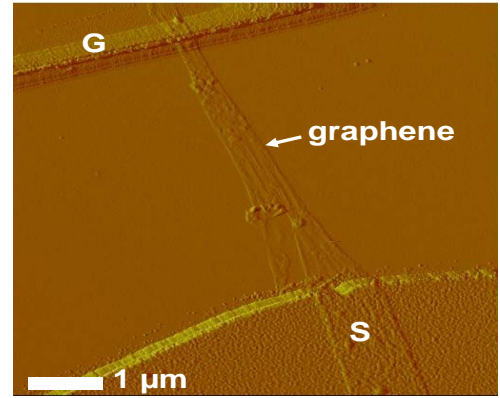

Fig. 4. Atomic force microscope image of a graphene FET.

For a detailed analysis, Raman spectroscopy has been used to distinguish between few- and monolayer graphene [11][12]. For comparison, the Raman-spectra of the graphene flake (open rectangles) and a few layer graphene sample (solid circles) are depicted in Fig. 5. The graphene monolayer is clearly identified by the shape and the position of the Raman spectrum [11].

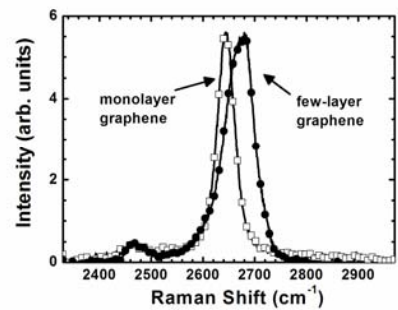

Fig. 5. Raman spectra of graphene flakes with different layer numbers.

## 3. Discussion

Electrical measurements have been performed on this monolayer graphene field-effect device prior and after top-gate deposition. Fig. 6 shows the output characteristics at zero gate voltage with and without a gate dielectric on top. The drain current increases linearly with increasing drain bias. No saturation is observed in the entire voltage range of $V_d = 0$ to 90 mV. After gate dielectric deposition the graphene FET shows a decreased drain current. Obviously, the presence of the gate dielectric on top of the graphene sheet has a large impact on the electrical characteristics of the graphene device.

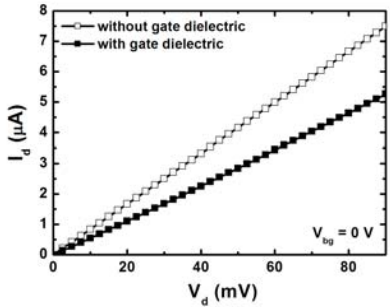

Fig. 6. Graphene FET output characteristics with and without top-gate dielectric at zero transversal field.

In Fig. 7, back-gate or pseudo-MOS transfer characteristics of a graphene FET before and after top-gate fabrication are shown. The source-drain voltage has been kept constant at $V_{ds} = 100$ mV and the back-gate electric field has been swept from $E_{bg} = -3.5$ MV/cm to $E_{bg} = 3.5$ MV/cm. The drain current is modulated by almost one order of magnitude without a top-gate (Fig. 7, black dots). Similar to CNTs, ambipolar behavior is observed. It is interesting to note that hole conduction is favored over electron conduction as negative back-gate fields result in higher drain current modulation compared to positive back-gate fields. This is attributed to unintentional chemical doping by adsorbants during processing and handling of the sample [13], which is also believed to shift the current minimum towards positive $E_{bg}$.

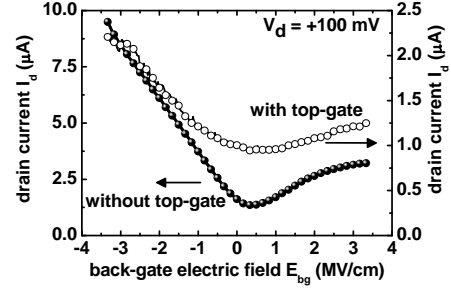

Fig. 7. Back-gate transfer characteristics of a graphene FET with and without top-gate.

After the deposition of the $SiO_2$ top-gate dielectric, the IV transfer characteristics maintain their basic signature, but the current level and the current modulation are decreased dramatically, represented by hollow circles in Fig. 7. This translates to a reduced carrier mobility in double gated graphene. This result seems to be in contrast with other groups which have reported no significant effect of top gate dielectrics on DC transport properties. Carrier mobilities have been calculated using a Drude model for carrier transport. First, the charge carrier density $n_s$ in the graphene FET in Fig. 7 has been calculated using

$$n_s = \varepsilon_{ox} * V_G / (q * t_{ox})$$

with the silicon dioxide permittivity $\varepsilon_{ox}$, gate voltage $V_G$, electron charge q and silicon dioxide thickness $t_{ox}$. The effective electric field in the graphene sheet has been calculated by using a value of $\varepsilon_g = 2.4$ for the dielectric constant of graphene [14].

For the uncovered graphene, mobilities of $\mu_h \geq 4790$ cm$^2$/Vs for holes and

$\mu_e \geq 4780$ cm$^2$/Vs for electrons at effective electric fields of $E_{eff} = 0.4$ MV/cm have been obtained [8].

Fig. 8 compares carrier mobilities in graphene with and without a top-gate electrode to the universal mobility of silicon [15] and to experimental literature data for ultra thin body SOI transistors [16] over a range of effective transversal electric field. A distinct mobility reduction after top-gate deposition is observed that is attributed to the participation of the top π-orbitales to van der Waals bonds to the silicon dioxide. The resultant reduction of orbital overlap then leads to a reduced conductivity [7].

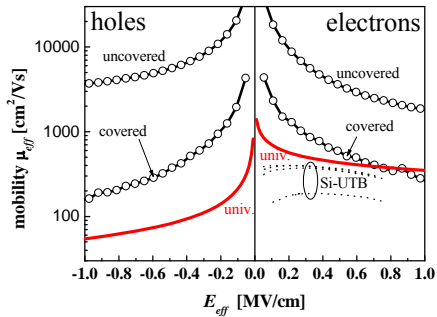

Fig. 8. Mobility curves of graphene FETs with and without top-gate.

Despite the substantial mobility reduction after the top-gate deposition, graphene mobility exceeds the universal mobility of silicon almost over the entire measured range. This is particularly encouraging since this result has been achieved with an evaporated top-gate oxide with a high charge trap density. Finally, a comparison with ultra-thin body SOI transistors further makes a good case for graphene. Here, literature reports electron mobilities below $\mu_e = 70$ cm$^2$/Vs in $t_{Si} = 2.5$ nm films [16] and hole mobility below $\mu_h \sim 60$ cm$^2$/Vs in 3.7 nm films [17][18] both in (100) silicon at room temperature.

The top-gate transfer characteristics of a graphene FET are shown in Fig. 9 for four different back-gate fields $E_{bg}$. The drain current $I_d$ is modulated by the top-gate field $E_{tg}$. The constant back-gate field $E_{bg}$ results in an offset of the top-gate transfer characteristics and does not change its ambipolar signature. The influence of the back-gate field is therefore mainly attributed to a modulation of the series resistance in the source and drain leads of the graphene FET.

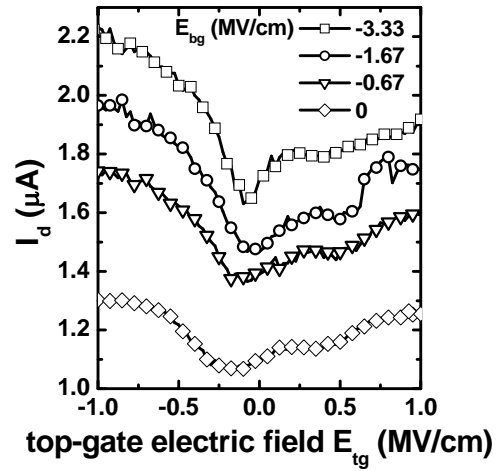

Fig. 9. Top-gate transfer characteristics of a Graphene FET for different back-gate fields $E_{bg}$.

## 4. Conclusion

Double-gated graphene field effect transistors are investigated in this work. In addition to previously investigated pseudo-MOS structures, a second gate is fabricated on top of graphene layers. IV-measurements show a reduction of drain currents caused by decreased carrier mobilities. Despite a poor quality top-gate oxide, however, graphene mobility values still exceed the universal mobility of silicon over almost the entire range investigated. Furthermore, graphene exceeds by far reported mobilities in UTB SOI MOSFETs. Even though band gap engineering has been theoretically proposed to improve graphene FET operation [19][20][21], this experimental work confirms the potential of graphene as a nanoelectronics material.

## 5. Acknowledgement

Financial support by the German Federal Ministry of Education and Research (BMBF) under contract number NKNF 03X5508 ("ALEGRA") is gratefully acknowledged.